\documentclass[10pt,letterpaper,twocolumn,english]{aastex6}
\usepackage{ae,aecompl}
\usepackage[T1]{fontenc}
\usepackage[latin9]{inputenc}
\usepackage{xcolor}
\usepackage{pdfcolmk}
\usepackage{array}
\usepackage{booktabs}
\usepackage{amstext}
\usepackage{graphicx}
\PassOptionsToPackage{normalem}{ulem}
\usepackage{ulem}

\makeatletter

\pdfpageheight\paperheight
\pdfpagewidth\paperwidth

\providecommand{\tabularnewline}{\\}
\newcommand{\lyxdot}{.}

\providecolor{lyxadded}{rgb}{0,0,1}
\providecolor{lyxdeleted}{rgb}{1,0,0}
\DeclareRobustCommand{\lyxsout}[1]{\ifx\\#1\else\sout{#1}\fi}

\makeatother

\usepackage{babel}
\begin{document}

\title{Non-Zeeman Circular Polarization of Molecular Spectral Lines in the
ISM}

\author{Mohammed Afif Chamma$^{1}$, Martin Houde$^{1}$, Josep Miquel Girart$^{2,3}$
and Ramprasad Rao$^{4}$}

\affil{$^{1}$Department of Physics and Astronomy, The University of Western
Ontario, London, ON, N6A 3K7, Canada}

\affil{$^{2}$Institut de Ciències de l'Espai (ICE, CSIC), Can Magrans,
S/N, E-08193 Cerdanyola del Vallès, Catalonia, Spain}

\affil{$^{3}$Institut d'Estudis Espacials de Catalunya (IEEC), E-08034,
Barcelona, Catalonia }

\affil{$^{4}$Submillimeter Array, Academia Sinica Institute of Astronomy
and Astrophysics, 645 N. Aohoku Place, Hilo, HI 96720, USA}
\begin{abstract}
Accurately measuring the magnetic field in the ISM is essential for
understanding star-formation processes. We searched archival data
of the Submillimeter Array (SMA) for evidence of circular polarization
in common molecular tracers, most notably CO. This circular polarization
possibly arises from anisotropic resonant scattering, which would
imply that some background linearly polarized radiation is being converted
to circular polarization. We find circular polarization in the star-forming
regions NGC7538 (in CO) and Orion KL (in CO and SiO), as well as in
the carbon star IRC+10216 (in CS, SiS, H$^{13}$CN and CO) at high
enough levels to suggest that the presence of circular polarization
in these spectral lines is common for such objects. This implies that
measuring circular polarization is important when studying magnetic
fields through the linear polarization of molecular spectral lines
in the interstellar medium. We also provide a simple Python wrapper
for the \emph{Miriad} data reduction package.
\end{abstract}


\received{May 14 2018}
\revised{July 11 2018}
\accepted{July 30 2018}
\slugcomment{Accepted to MNRAS}

\section{Introduction}

\label{sec:Introduction}Understanding the role of magnetic fields
in star-forming regions allows us to test ideas about free-fall collapse
and support mechanisms in molecular clouds, filling in details about
the star formation process. By measuring the radiation from star-forming
regions astronomers use polarimetry to infer the magnitude and orientation
of the magnetic field. The Davis-Chandrasekhar-Fermi (DCF) method
\citep{Davis1951,Chandrasekhar1953} uses the dispersion of polarization
angles (PA) of linear polarization (LP) for measuring the plane-of-the-sky
component of the magnetic field. The presence of a magnetic field
leads to LP radiation because dust and molecules will align themselves
relative to the field. Aligned particles can emit radiation with a
net level of LP greater than zero. Aligned dust can also absorb radiation
whose polarization is aligned with its long axis, acting as a sort
of polarizing grid. Thus measuring the amount of dust LP in the infrared
continuum tells us about the degree to which the dust is aligned with
the magnetic field, which in turn can tell us about the strength of
the magnetic field \citep{Davis1951,Chandrasekhar1953,Crutcher2012}. 

The alignment of molecules and their interaction with the ambient
magnetic field can cause their transitions to be linearly polarized
by a few percent through the so-called Goldreich-Kylafis (GK) effect
\citep{Goldreich1981}. The GK effect can occur for molecular lines
with even weak Zeeman splitting when the radiation field (for example)
is anisotropic and the splitting frequency is greater than the collisional
rate and other radiative processes. These conditions can cause a population
imbalance in the magnetic sublevels that results in a net LP aligned
either perpendicular or parallel to the plane-of-the-sky component
of the magnetic field. LP due to the GK effect was first detected
in CS ($J=2\rightarrow1)$ and HCN $(J=1\rightarrow0$) in IRC+10216
and CRL 2688 by \citet{Glenn1997} and later in CO ($J=3\rightarrow2$)
and ($J=2\rightarrow1)$ by \citet{Greaves1999}. Shortly after, LP
from the GK effect was detected interferometrically for the first
time and used to map the direction of the magnetic field in NGC1333
IRAS4A \citep{Girart1999}. As with dust, the PA associated with the
LP in the spectral line can be measured and used to infer properties
of the magnetic field through a dispersion analysis \citep{Davis1951,Chandrasekhar1953,Crutcher2012}.
Observations in DR 21(OH) of LP in CO $(J=2\rightarrow1)$ were compared
with simultaneous dust continuum polarization measurements to assess
the accuracy in using the GK effect to probe the magnetic field \citep{Lai2003,Cortes2005}.

For molecular lines circular polarization (CP) is usually ignored,
largely because of difficulty in its measurement and its assumed irrelevance.
However, a significant amount of unexpected circular polarization
was reported by \citet{Houde2013} in CO $(J=2\rightarrow1)$, a common
tracer through LP of magnetic fields \citep{Crutcher2012}, using
FSPPol \citep{Hezareh2010} at the Caltech Submillimeter Observatory
(CSO). The presence of CP in a molecular transition can be explained
with Zeeman splitting for some molecules/transitions possessing a
significant magnetic moment (e.g., CN), but CO is highly insensitive
to the Zeeman effect. In addition, the observed Stokes $V$ profile
in Orion KL was positive and symmetric, which is also unexpected since
Zeeman splitting often gives rise to an approximately antisymmetric
Stokes $V$ profile. To explain this detection a model was proposed
whereby background LP radiation is converted to CP radiation through
anisotropic resonant scattering (ARS) \citep{Houde2013,Houde2014}.
This was further tested in the supernova remnant IC443 by \citet{Hezareh2013},
where the measured CP flux of CO lines ($J=2\rightarrow1$) and ($J=1\rightarrow0$)
were `re-inserted' into the measured LP flux to account for the LP-to-CP
conversion and its effect on the PA. They found that the PAs obtained
from the different CO transitions only agreed with each other and
with those obtained from dust polarimetry after the CP flux was accounted
for. If ARS is common to other objects then using LP in CO as a tracer
of the magnetic field will introduce a systematic error unless the
CP of CO lines is also measured. 

The main goal of this paper is to find further evidence of CP in more
objects and molecular lines through a search of archival data of the
Submillimeter Array (SMA). In Section \ref{sec:Measurement-with-Radio}
we discuss the issues that arise when doing polarimetry with radio
interferometry, focusing specifically on circular polarization (CP).
Section \ref{sec:Squint-Correction} will give in detail our scheme
for correcting a spurious source of CP that arises with the SMA. Section
\ref{sec:Observations} presents archival observations of four objects
made with the SMA on Mauna Kea and investigates the reliability of
our CP detections. Finally, in Section \ref{sec:Discussion} we will
highlight the significance of these CP detections and summarize their
implications.

\section{Measurement of CP with Radio Interferometry}

\label{sec:Measurement-with-Radio}The measurement of CP is challenging
to calibrate, especially when using radio interferometers like the
SMA or ALMA. The SMA, which has linear feeds, uses a quarter-waveplate
(QWP) in the path of the antenna beam to convert incident LP light
to CP light to measure LP signals before sending them to the correlator.
While not its most common use at the SMA the QWP can work the other
way to measure CP: incident CP is converted to LP and then measured
by the receivers. On the other hand, ALMA which also has linear feeds
does not use a QWP and measures LP signals directly. While both measurement
techniques can be used to measure CP, the calibration process is different
\citep{Sault1996}. Despite the calibration challenges the SMA has
been used to take precise measurements of CP from dust continuum in
Sgr A{*} \citep{Munoz2012}. Additionally, the VLBA, equipped with
circular feeds, has been used to take full Stokes $I$, $Q$, $U$,
and $V$ measurements of SiO masers at 43.1 and 42.8 GHz \citep{Cotton2011}.
In that work a correction for spurious CP analogous to the correction
detailed in Section \ref{sec:Squint-Correction} is performed for
accurate calibration of Stokes $V$.

For a discussion on measuring CP with radio interferometry and on
design choices at the SMA (such as the choice of converting from LP-to-CP
and vice-versa) see \citet{Hamaker1996} and \citet{Marrone2008}.

\subsection{Linear vs. Circular Feeds}

To illustrate briefly the differences between the two feed types consider
the following: with an orthogonal CP basis, the Stokes $V$ parameter
for a beam of radiation is defined by $V\propto\langle E_{R}^{2}\rangle-\langle E_{L}^{2}\rangle$,
where $E_{R}$ and $E_{L}$ are the right-CP and left-CP electric
fields, respectively. With an orthogonal LP basis Stokes $V$ is defined
by $V\propto-2\,\text{Im}(E_{x}E_{y}^{*})$, where $E_{x}$ and $E_{y}$
are the LP fields, and $\text{Im}()$ denotes the imaginary part.
In the CP basis case we take the difference of two measured intensities,
while the linear feed case requires us to measure the phase of the
electromagnetic wave. 

Now, when the measurement is made with interferometry it is the visibilities
\textendash{} the correlated waveforms between a pair of antennae
\textendash{} that are measured. In the CP basis case the Stokes $V$
visibility scales as $\mathcal{V}_{\text{V}}\propto\text{\ensuremath{\mathcal{V}}}_{RR}-\text{\ensuremath{\mathcal{V}}}_{LL}$,
where $\text{\ensuremath{\mathcal{V}}}_{RR}$ and $\text{\ensuremath{\mathcal{V}}}_{LL}$
are the visibilities obtained from correlating two antennae measuring
right-CP and left-CP, respectively (see eq. (\ref{eq:CPV}) below).
In the linear feed case the Stokes $V$ visibility is coupled with
the Stokes $Q$ and $U$ visibilities \citep[see Section 4.1 of][]{Thompson2001}.
This means the Stokes $Q$ and $U$ of any calibration object must
be measured as well. This is not possible with the SMA setup used
for the observations discussed here.

The SMA uses observations of a bright point source over a large range
of parallactic angles to distinguish between source and instrumental
polarization and determine the polarization leakage terms. Determining
these leakage terms is crucial for accurate Stokes $Q$ and $U$ measurements,
however Stokes $I$ and $V$ are independent of the leakage terms
to first-order \citep{Thompson2001,Marrone2008,Munoz2012}. 

\section{Squint Correction}

\label{sec:Squint-Correction}We now describe spurious Stokes $V$
signals that can arise during observations such as those presented
here and our scheme for correcting them. This instrumental Stokes
$V$ comes from a slight pointing offset between the left- and right-handed
CP beams.

The archival data used were in all cases observed with the goal of
measuring LP (i.e., the Stokes $Q$ and Stokes $U$ parameters). As
mentioned earlier on the SMA this is done with a QWP placed in front
of the LP receivers to convert incident CP (LP) to LP (CP). While
this method suffers from the errors that arise when subtracting two
large measurements from each other, when performing CP measurements,
it avoids having to solve for the LP terms of calibration objects
\citep{Marrone2008,Thompson2001}. 

The output of the interferometer is the visibility, which consists
of the cross-correlation between the voltage signals from a pair of
antennae and can be written as 
\begin{equation}
\mathcal{V}(u,v)=\frac{\langle V_{a}(t)\star V_{b}(t)\rangle}{A_{0}\Delta\nu},
\end{equation}
where $A_{0}$ is the collecting area of the antennae, $\Delta\nu$
is the bandwidth, and $\langle V_{a}(t)\star V_{b}(t)\rangle$ is
the time-averaged cross-correlation of voltage signals $V_{a}(t)$
and $V_{b}(t)$ from antennae $a$ and $b$, respectively. The arguments
$u$ and $v$ are determined by the baseline separation and orientation
of the two antennae. Thus a pair of antennae samples a single point
of the visibility function $\mathcal{V}(u,v)$ \citep{Thompson2001}.

Through the van Cittert-Zernike theorem it can be shown that the visibility
function $\mathcal{V}(u,v)$ gives the Fourier transform of the source
intensity $I(l,m)$ for angular position $l$ and $m$ \citep[chap. 3 of][]{Thompson2001}.
Thus we have 
\begin{equation}
\mathcal{V}(u,v)=\int\int e^{-iul}e^{-ivm}I(l,\,m)\,dl\,dm.\label{eq:FT}
\end{equation}

In practice the true source intensity $I(l,m)$ is convolved with
the instrumental response to a point source, called the instrumental
beam, and a deconvolution step is necessary to obtain the true map.

\begin{figure}
\begin{centering}
\includegraphics[angle=270,scale=0.45]{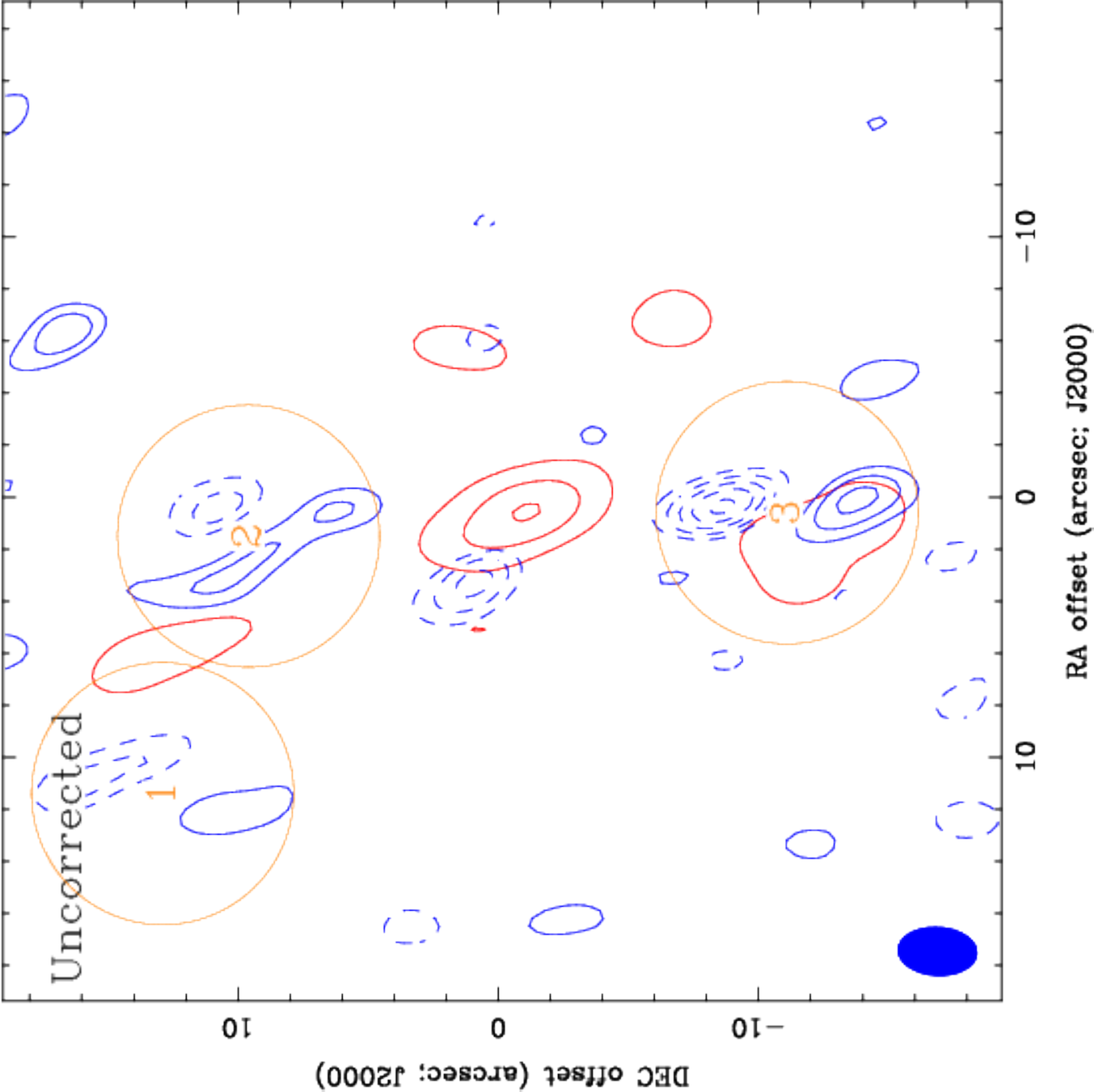}
\par\end{centering}
\begin{centering}
\includegraphics[angle=270,scale=0.45]{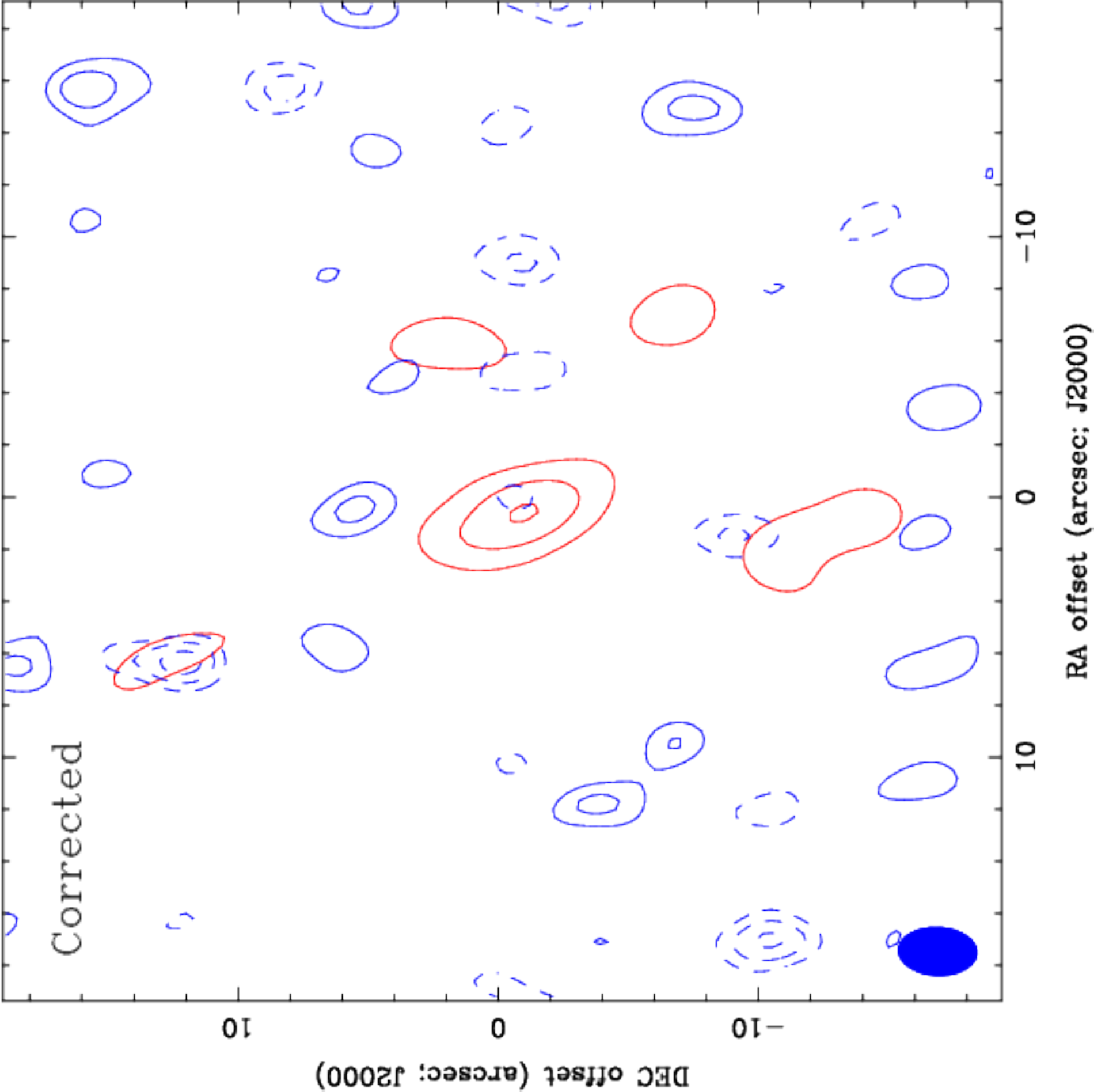}
\par\end{centering}
\caption{\label{fig:beforeafter}Map of the continuum around 345GHz in Orion
KL before (top) and after (bottom) squint correction. Red contours
are Stokes $I$, blue contours are Stokes $V$. Dashed lines denote
negative values, solid lines denote positive values. Note the three
pairs of positive and negative Stokes $V$ peaks in the uncorrected
map (identified and circled in orange). These largely disappear after
correction. Red Stokes $I$ contours are at 15\%, 50\%, and 95\% of
the peak intensity. Blue Stokes $V$ contours are at -8, -7, -6, -5,
-4, -3, -2, 2, 3, 4, 5, 6, 7 and 8$\sigma$ levels. }
\end{figure}

Given antennae $a$ and $b$, the Stokes $V$ visibility in the circular
feed case is given by \citep{Thompson2001,Munoz2012}
\begin{equation}
\mathcal{V}_{V}\simeq\frac{1}{2}\Big\{\mathcal{V}_{RR}/(g_{Ra}g_{Rb}^{*})-\mathcal{V}_{LL}/(g_{La}g_{Lb}^{*})\Big\},\label{eq:CPV}
\end{equation}
where $\mathcal{V}_{RR}$ and $\mathcal{V}_{LL}$ are the right-handed
CP and left-handed CP visibilities, respectively, measured by appropriately
orienting the QWP that is placed in the beam of the antennae and correlating
their responses \citep{Marrone2008}. The complex gain factors for
each polarization for each antenna are $g_{Ra}$, $g_{Rb}$, $g_{La}$
and $g_{Lb}$ where $R$ and $L$ are for right- and left-CP, respectively.
Because the Stokes $V$ visibility is found by taking the difference
of two beams a slight offset gives rise to pairs of positive and negative
peaks of Stokes $V$, as can be seen in Figure \ref{fig:beforeafter},
for example. The cause of this offset is uncertain, but possibly arises
due to slight differences in the index of refraction of the QWP when
it is rotated.

To correct this offset we first note that since the visibilities are
the Fourier transform of the intensity map (eq. \ref{eq:FT}), an
offset in image space results in a complex factor in visibility space
that can be absorbed into the gain coefficients. To see this, consider
a map $I(l,m)$ that represents the true intensity $I$ at angular
position $l$ and $m$. If the instrument introduces an arbitrary
offset to position $(l_{0},m_{0})$, then the final image we calculate
is shifted such that $I'(l,m)=I(l-l_{0},m-m_{0})$ and the measured
visibility of the shifted map becomes 
\begin{eqnarray}
\mathcal{V}'(u,v) & = & \int\int e^{-iul}e^{-ivm}I(l-l_{0},m-m_{0})\,dl\,dm\nonumber \\
 & = & \int\int e^{-iu(\alpha+l_{0})}e^{-iv(\delta+m_{0})}I(\alpha,\delta)\,d\alpha\,d\delta\nonumber \\
 & = & e^{-iul_{0}}e^{-ivm_{0}}\int\int e^{-iu\alpha}e^{-iv\delta}I(\alpha,\delta)\,d\alpha\,d\delta\nonumber \\
 & = & e^{-iul_{0}}e^{-ivm_{0}}\mathcal{V}(u,v)\nonumber \\
 & \equiv & g_{\text{offset}}\mathcal{V}(u,v),
\end{eqnarray}
where we used the change of variables $\alpha=l-l_{0}$ and $\delta=m-m_{0}$.
We thus find the aforementioned complex factor that multiplies the
true visibility $\mathcal{V}(u,v)$. It is therefore easiest to correct
for the offset in visibility-space by using \emph{Miriad }\citep{Sault1995}
to solve for the gain coefficients on each of the two polarized beams,
independently of each other. Specifically the process is:
\begin{enumerate}
\item The observations are calibrated for gain and phase in the usual way
using observations of known sources (usually quasars like 3C84, 3C454,
etc.; see Miriad User Guide \citep{Sault2008}).
\item The visibilities are split into line-free continuum and line data
that are then mapped to obtain models using the `\texttt{clean}' algorithm
\citep{Sault2008}.
\item The separate continuum and line data are further split into \emph{LL}
and \emph{RR} visibilities. \emph{Miriad'}s \texttt{selfcal} is used
on the continuum data to solve for the gain coefficients of each antenna
and each polarization (\emph{L} or \emph{R}). This is done by minimizing
the difference between measured visibilities $\mathcal{V}_{ij}$ of
antennae $i$ and $j$ and model visibilities $\hat{\mathcal{V}}_{ij}$
according to $\epsilon^{2}=\sum|\mathcal{V}_{ij}-g_{i}g_{j}^{*}\hat{\mathcal{V}}_{ij}|^{2}$
for each of the correlations \emph{LL} and \emph{RR} \citep{Schwab1980,Sault2008}.
The model visibilities used are those found earlier. Note the subscripts
here denote a specific antenna and not the polarization state as before. 
\item The gain found from the continuum \emph{LL} data is then applied to
the line \emph{LL} data, and similarly for the \emph{RR} continuum
and line data.
\item The different visibilities (\emph{LL}, \emph{RR}, \emph{RL}, \emph{LR})
are recombined and inverted to produce corrected and CLEANed maps.
Spectra can be obtained either from the corrected visibilities or
the maps. 
\end{enumerate}
Note that in the deconvolution steps (steps 2 and 5) the same instrumental
beam is used for all polarizations. This is due to the behaviour of
\emph{Miriad}'s \texttt{invert}, which produces a single instrumental
beam corresponding to all image planes and Stokes parameters \citep[sec. 13.4 of][]{Sault2008}.
We checked that the instrumental response to different polarizations
was similar by inverting the $LL$- and $RR$-handed visibilities
separately from one other and finding that the instrumental beams
for both polarizations to be almost identical.

Figure \ref{fig:beforeafter} shows maps before- and after- correction
maps for the continuum in Orion KL around 345GHz. We see that before
correction there are large peaks of Stokes $V$, and that the three
(identified) positive peaks have a negative peak close by. After the
correction, though there are still noisy Stokes $V$ signals throughout
the image, the pairs of peaks have disappeared.

\section{Observations}

\label{sec:Observations}We collected radio interferometric polarimetry
observations from the Submillimeter Array (SMA) archive that had been
measured using the linear-to-circular QWP equipment, a similar setup
to that used by \citet{Munoz2012} to measure circular polarization
from dust continuum in Sgr A{*}. Because the archival observations
were not taken with measurements of CP in mind the SNR is often low,
and we had to average velocity channels to increase it at the cost
of spectral resolution. This generally increases the SNR from 3-4
to 6-10. 

The four objects we present here are Orion KL, NGC7538, NGC1333 IRAS2A
and IRC+10216. The first three are well-known star-forming regions,
while IRC+10216 is an evolved carbon star. The data for Orion KL were
previously used for a dust polarization study in \citet{Tang2010},
and the data for IRC+10216 were previously used for spectral line
polarimetry in \citet{Girart2012}. The archival data for NGC1333
IRAS2A and NGC7538 used here have not been published before as far
as we are aware. We find significant Stokes $V$ signals in all objects
except for NGC1333 IRAS2A. Table \ref{tab:Summary-of-Archival} shows
a summary of the objects presented and related information.

\begin{table}
\begin{centering}
\begin{tabular}{>{\raggedright}p{0.19\columnwidth}>{\raggedright}p{0.33\columnwidth}>{\raggedleft}p{0.15\columnwidth}>{\raggedleft}p{0.25\columnwidth}}
\toprule 
\textbf{\footnotesize{}Object} & \textbf{\footnotesize{}Coordinates (J2000)} & \textbf{\footnotesize{}Array} & \textbf{\footnotesize{}Obs. Date}\tabularnewline
\midrule
\midrule 
\textbf{Orion KL} & RA 05$^{\text{h}}$35$^{\text{m}}$14.501$^{\text{s}}$\\
Dec -05$^{\circ}$22'30.40'' & Compact & 2008-01-06\tabularnewline
\midrule 
\textbf{NGC7538} & RA 23$^{\text{h}}$13$^{\text{m}}$44.771$^{\text{s}}$\\
Dec +61$^{\circ}$26'48.85'' & Compact & 2014-10-28\tabularnewline
\midrule 
\textbf{IRC+10216} & RA: 09$^{\text{h}}$47$^{\text{m}}$57.381$^{\text{s}}$\\
Dec +13$^{\circ}$16'43.70'' & Compact & 2009-11-24\tabularnewline
\midrule 
\textbf{NGC1333} & RA 03$^{\text{h}}$28$^{\text{m}}$55.580$^{\text{s}}$\\
Dec +31$^{\circ}$14'37.10'' & Compact & 2010-10-14\tabularnewline
\bottomrule
\end{tabular}
\par\end{centering}
\caption{\label{tab:Summary-of-Archival}Summary of Archival Observations Used}
\end{table}

The visibility data were corrected for beam squint, as explained in
Section \ref{sec:Squint-Correction} in order to reduce spurious Stokes
$V$ signals. As previously mentioned squint typically causes distinct
pairs of positive and negative peaks of Stokes $V$ throughout the
inverted image. The squint correction is confirmed visually by inspecting
the Stokes $V$ maps, where we see the pairs of peaks largely disappear.

\begin{figure*}
\vspace{-1cm}
\begin{centering}
\includegraphics[scale=0.35]{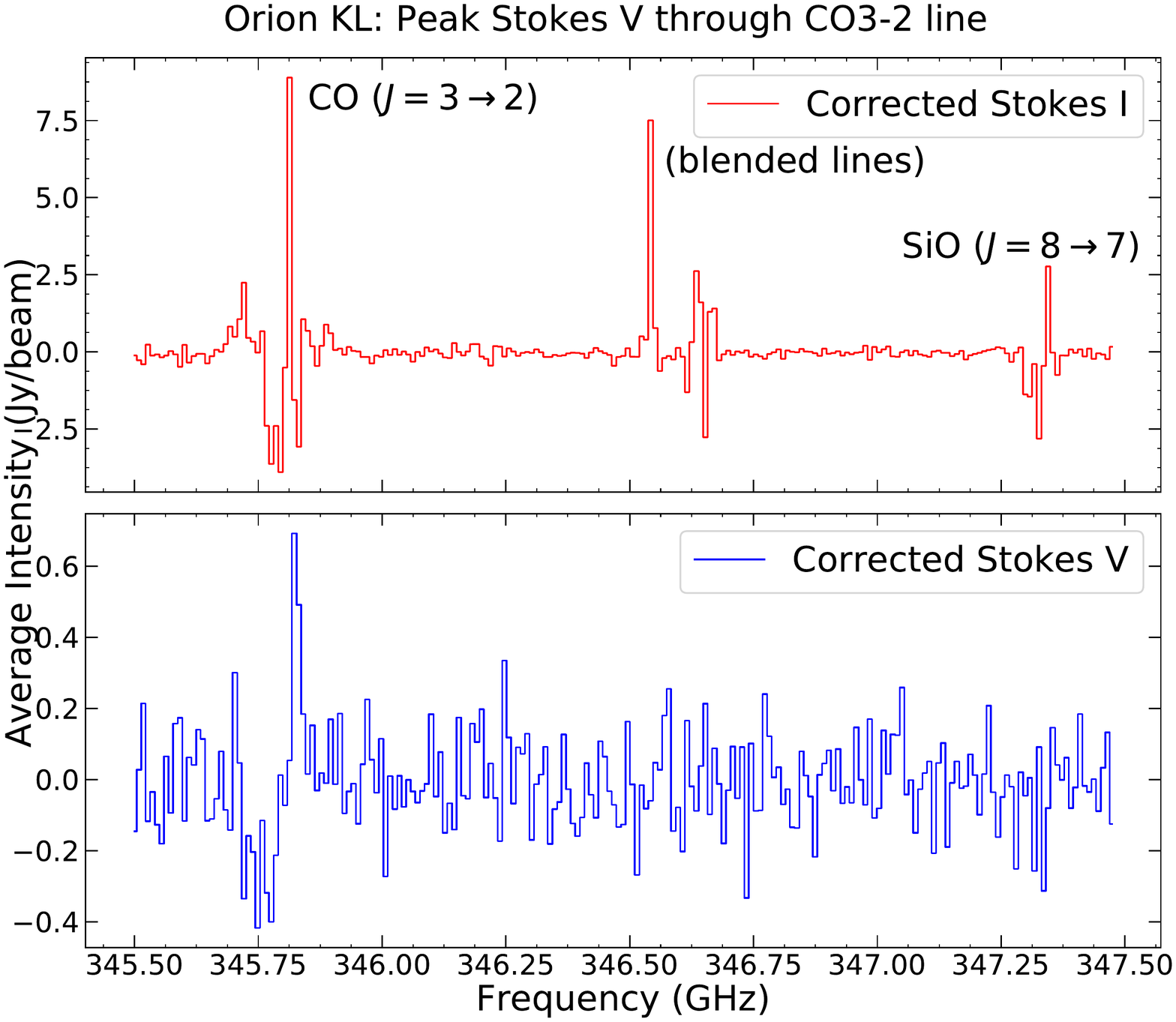}\includegraphics[scale=0.37,angle=270,origin=c]{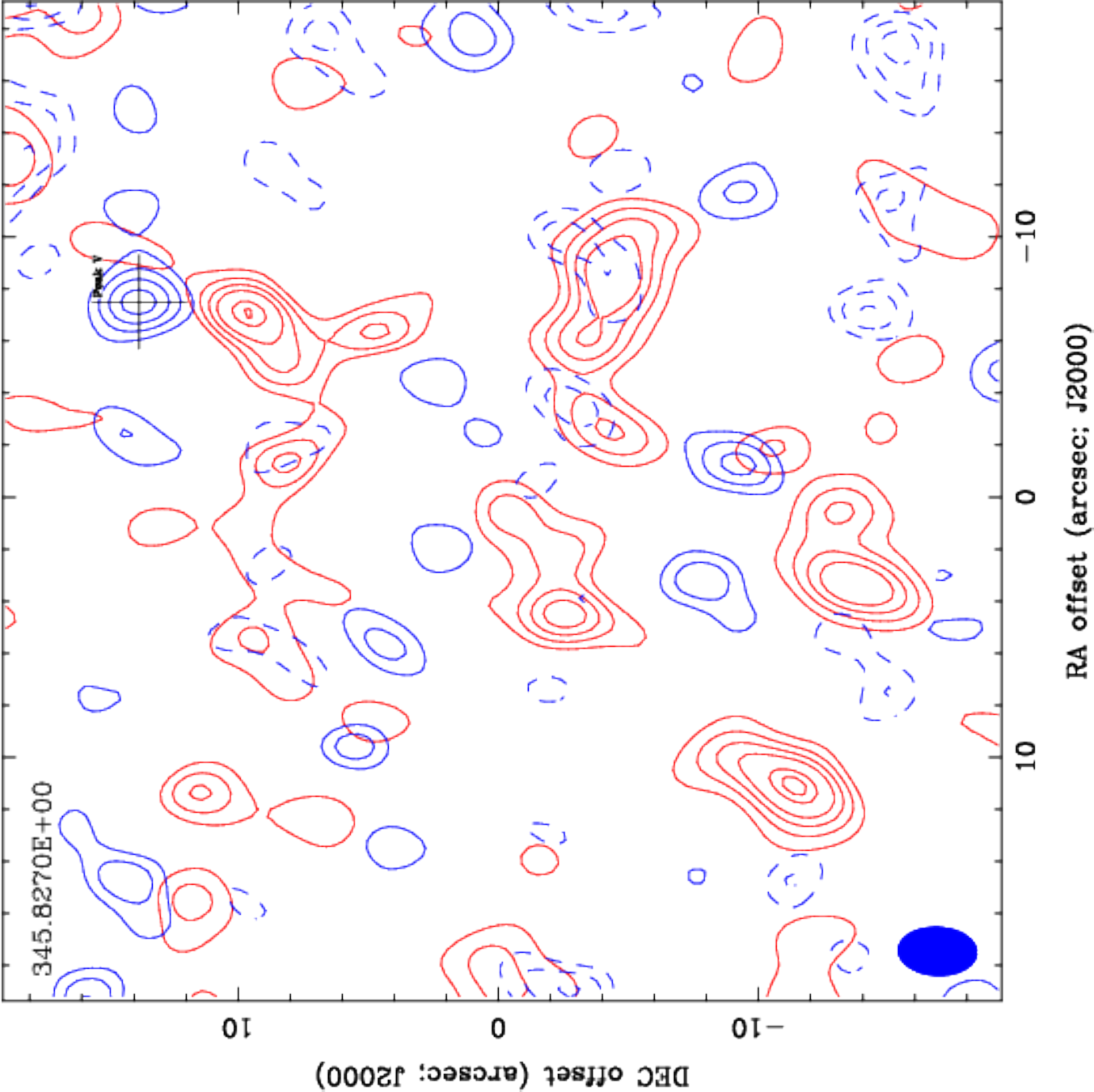}
\par\end{centering}
\begin{centering}
\includegraphics[scale=0.35]{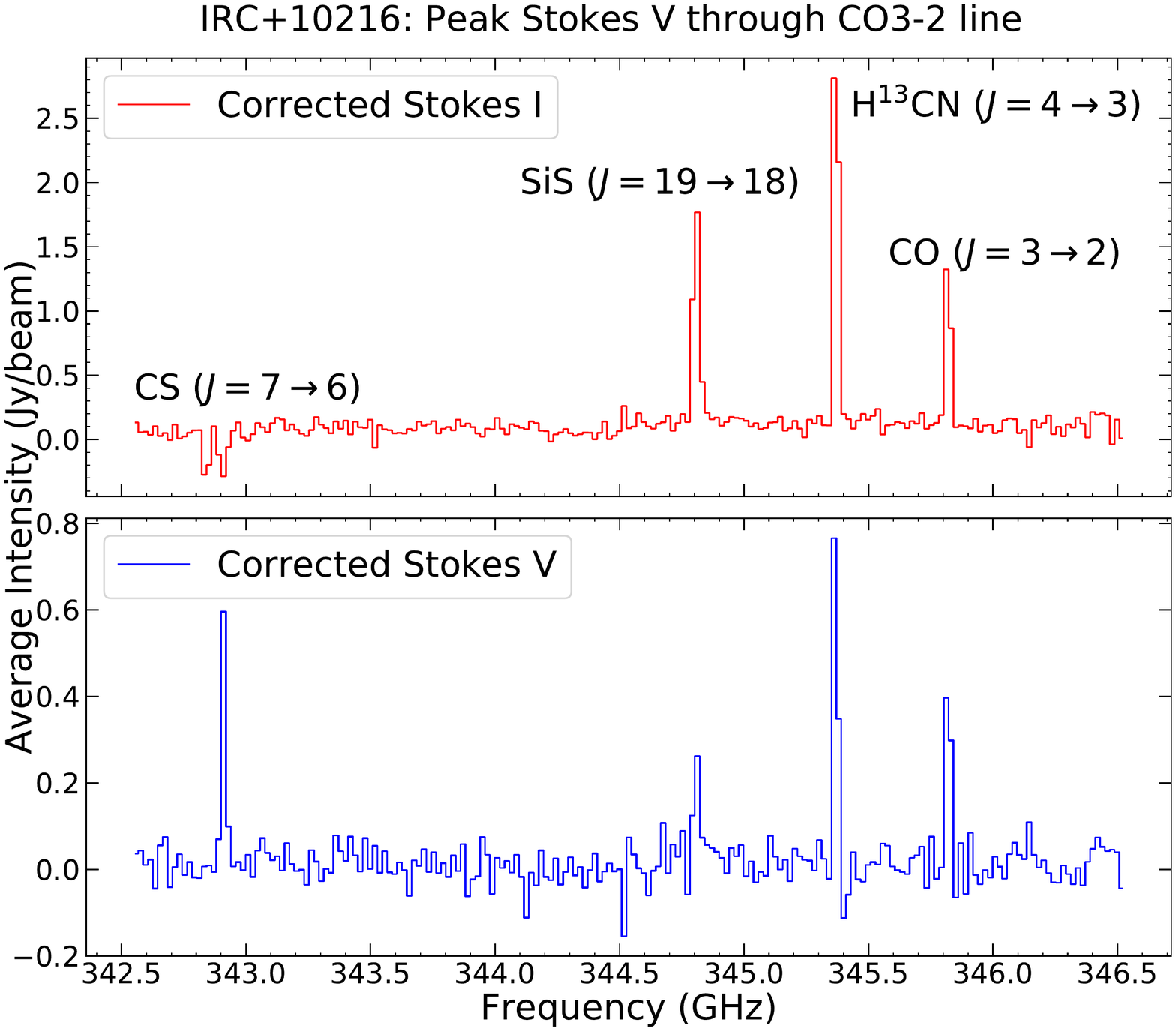}\includegraphics[angle=270,origin=c,scale=0.37]{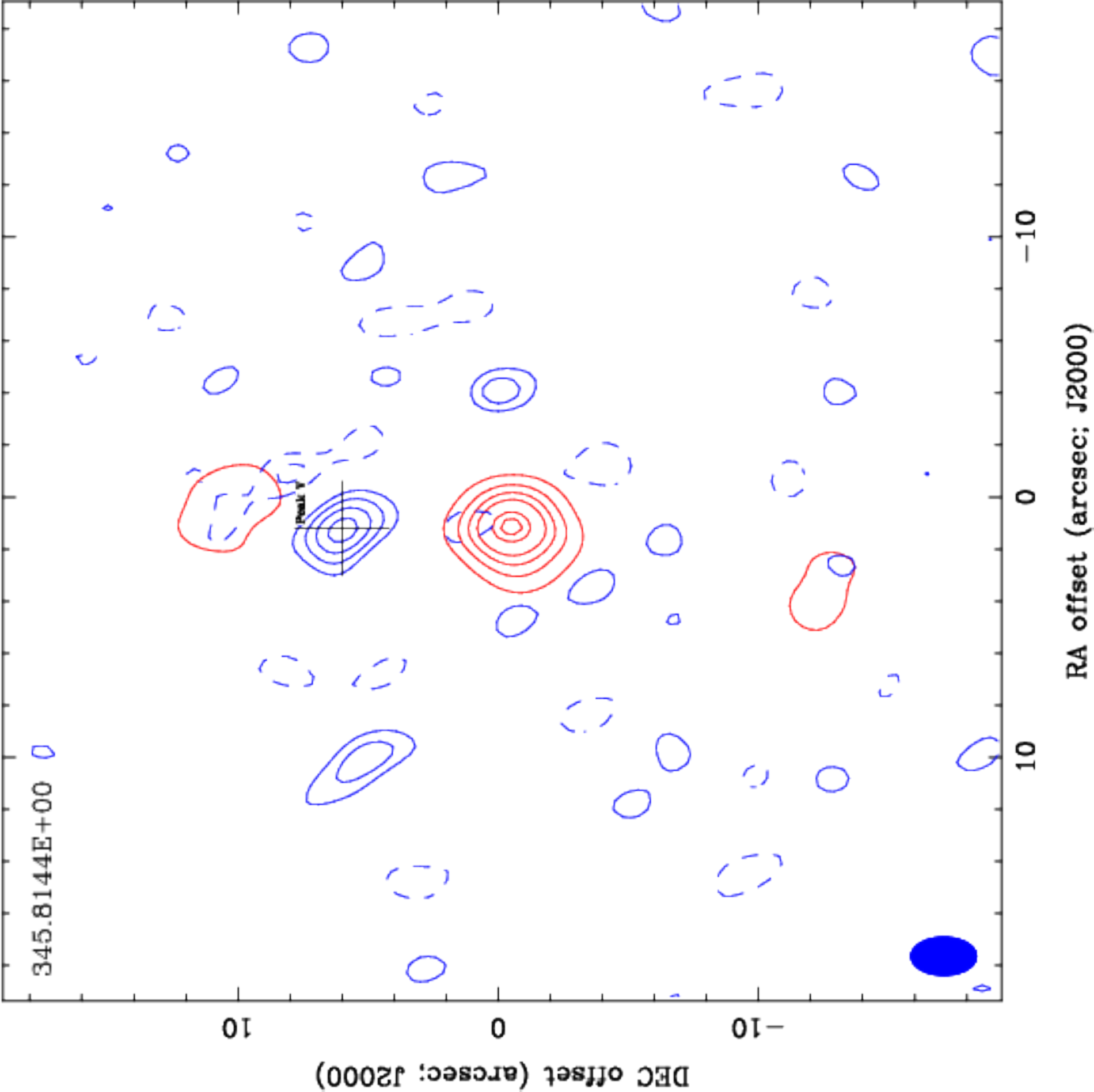}
\par\end{centering}
\caption{\label{fig:results}Corrected spectra and maps of the CO ($J=3\rightarrow2$)
line (345.8GHz) for Orion KL (top) and IRC+10216 (bottom). \textbf{\emph{Spectra}}\textbf{:}
\emph{Miriad}'s \texttt{maxfit} is used on the CO map to obtain the
location on the image where the Stokes $V$ signal at 345.8GHz is
maximum, and a spectrum is obtained through that point. The cross
on the map denotes the location of the Stokes $V$ peak. The red line
is Stokes $I$ and the blue is Stokes $V$.\textbf{\emph{ Maps}}\textbf{:}
Blue contours are Stokes $V$ and are shown at the -4, -3, -2, 2,
3, 4$\sigma$ levels. The RMS error for each Stokes $V$ map is found
using \emph{Miriad'}s \texttt{imstat} command: $\sigma=0.30$ and
0.17 Jy/beam, for Orion KL and IRC+10216, respectively. The distance
of the peak to the phase center is 15'' and 6'', respectively. Dark
red contours are Stokes $I$ and the levels are 15\%, 30\%, 45\%,
60\%, 85\% and 95\% of the maximum. The value in the top left is the
central frequency in GHz of the mapped signal, integrated over a narrow
bandwidth of $\sim$2 MHz. }
\end{figure*}

\begin{figure*}
\vspace{-1cm}
\begin{centering}
\includegraphics[scale=0.35]{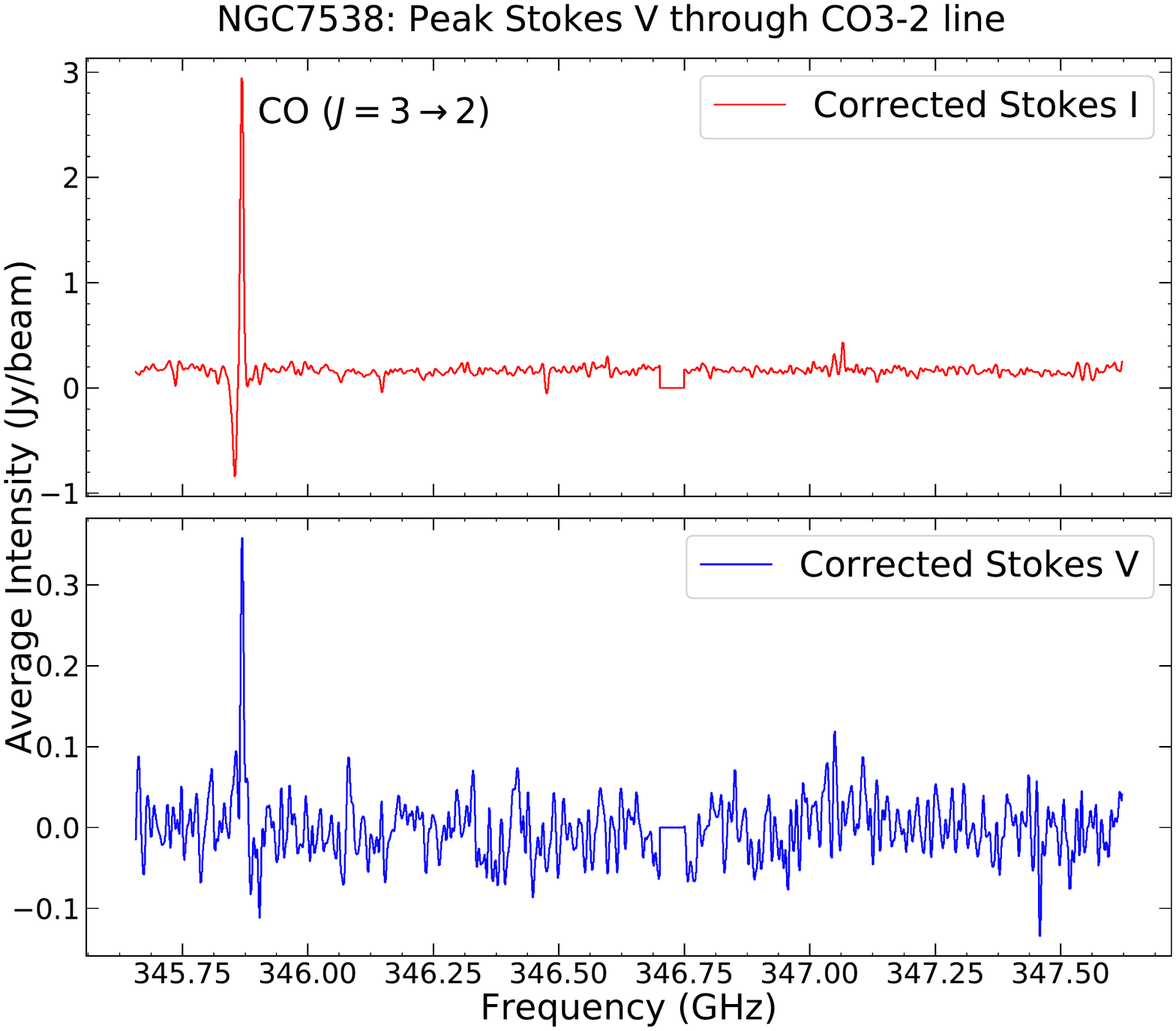}\includegraphics[scale=0.37,angle=270,origin=c]{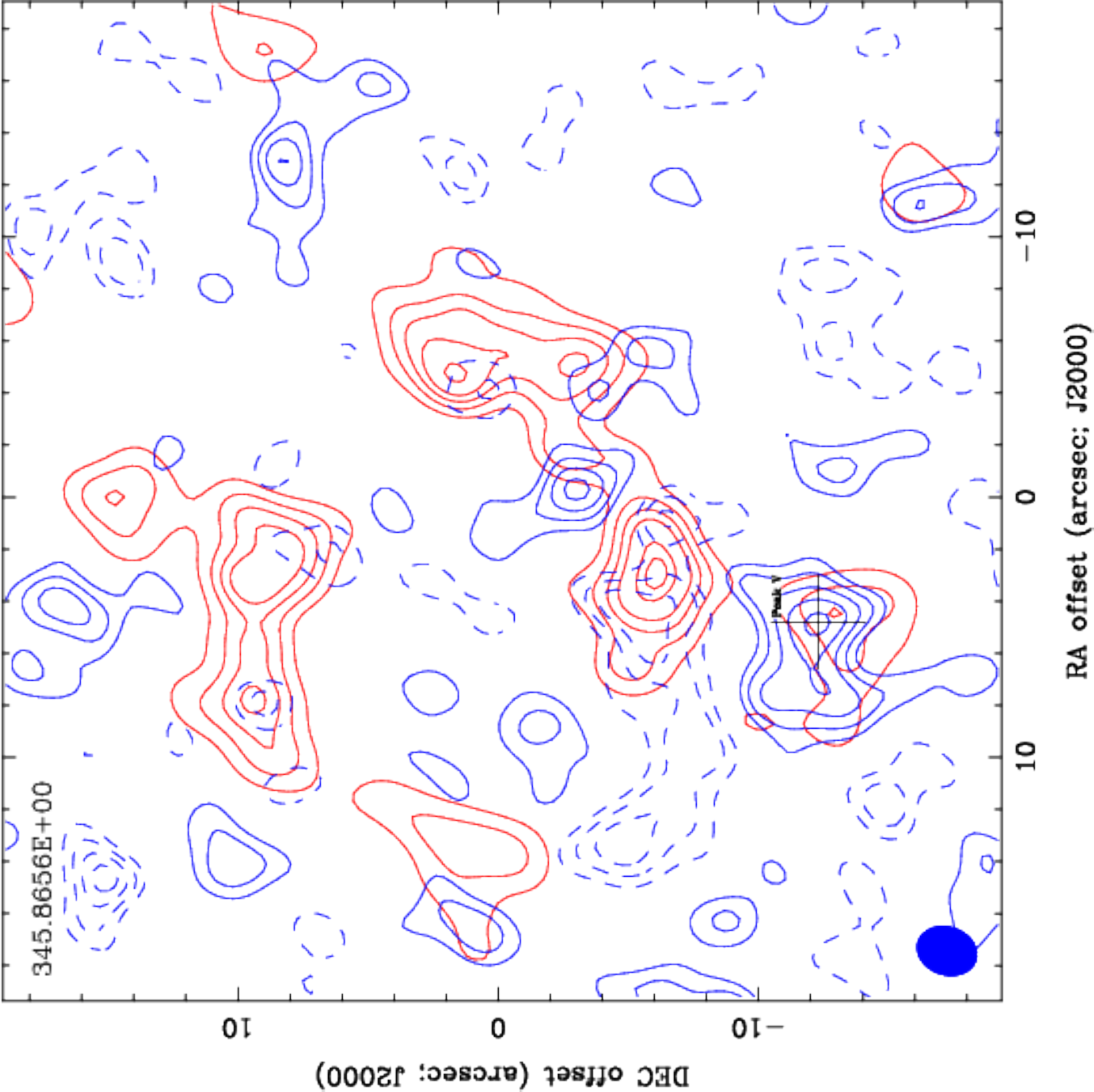}
\par\end{centering}
\begin{centering}
\includegraphics[scale=0.35]{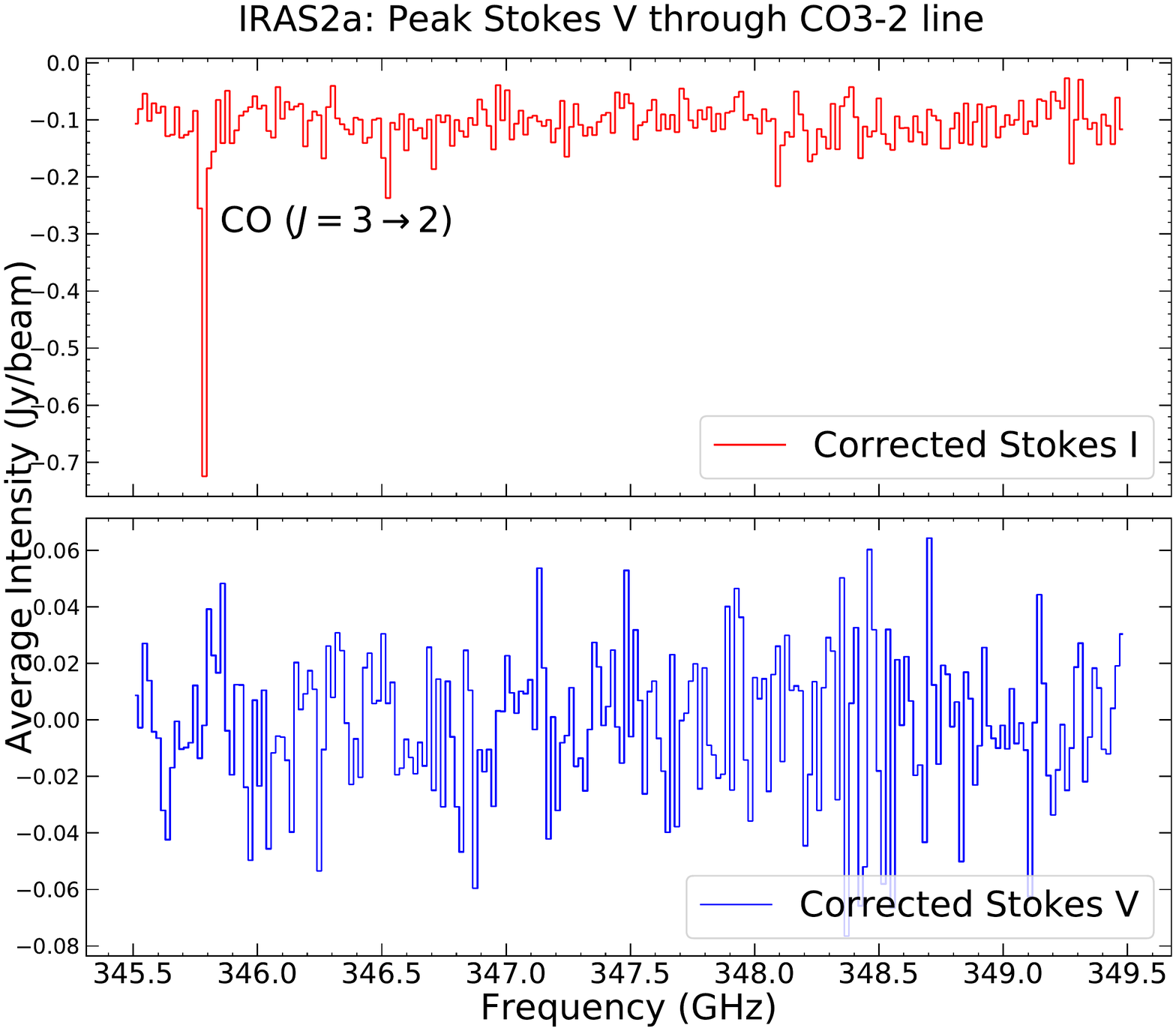}\includegraphics[angle=270,origin=c,scale=0.37]{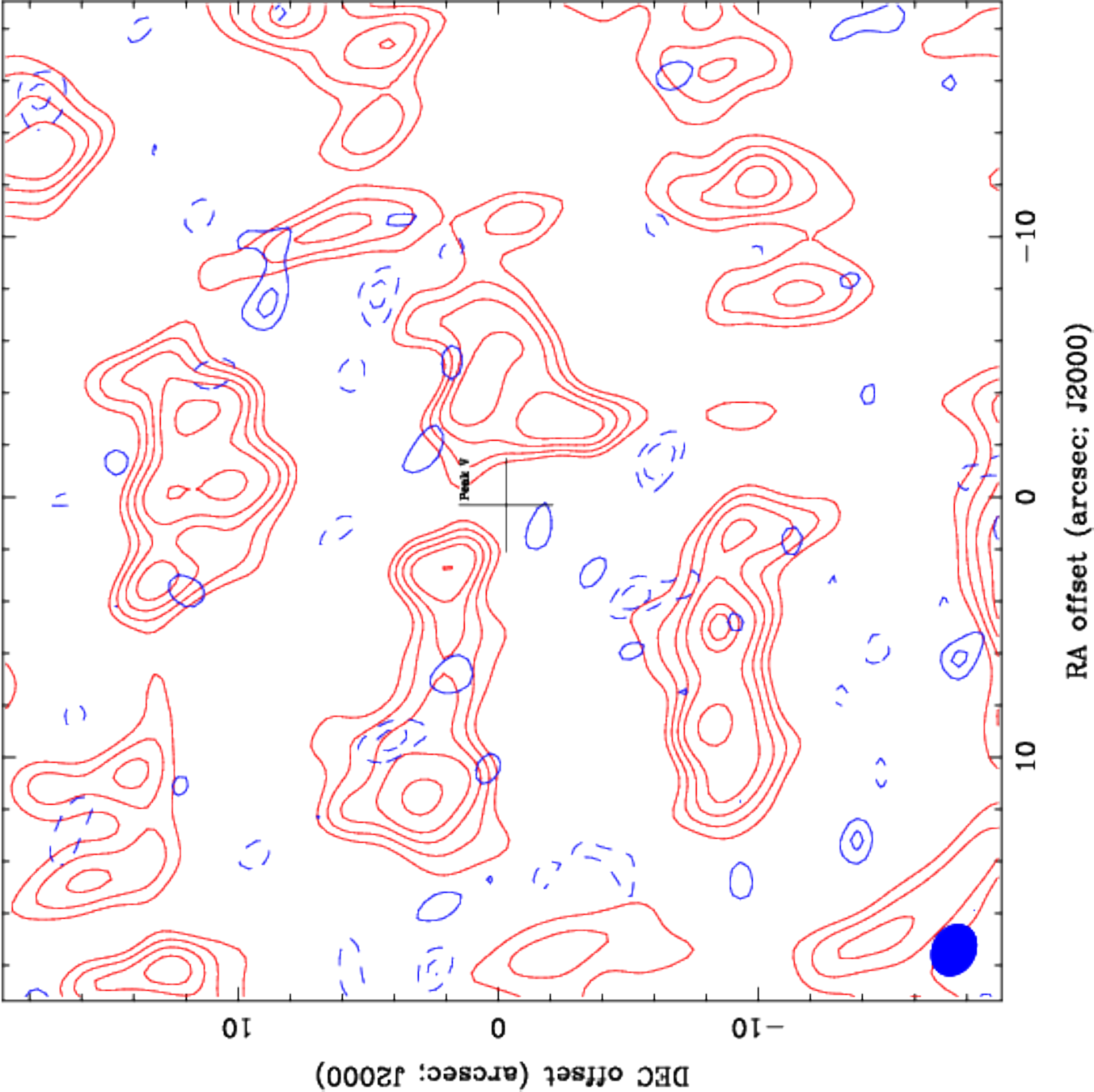}
\par\end{centering}
\caption{\label{fig:iras2a}Same as Figure \ref{fig:results} but for NGC7538
(top) and NGC1333 IRAS2A (bottom). The spectrum for NGC7538 is Hanning
smoothed. No significant Stokes $V$ signal is detected in NGC1333
IRAS2A, probably because the object is too dim. Contours are the same
levels as in Figure \ref{fig:results} and the spectrum is obtained
the same way. The RMS for the NGC7538 and NGC1333 IRAS2A maps is $\sigma=0.15$
and $0.12$ Jy/beam, respectively. The distance of the peak to the
phase center is 13'' for NGC7538.}
\end{figure*}

\begin{figure*}
\vspace{-0.5cm}
\begin{centering}
\includegraphics[bb=20bp 40bp 565bp 780bp,clip,width=0.8\paperwidth]{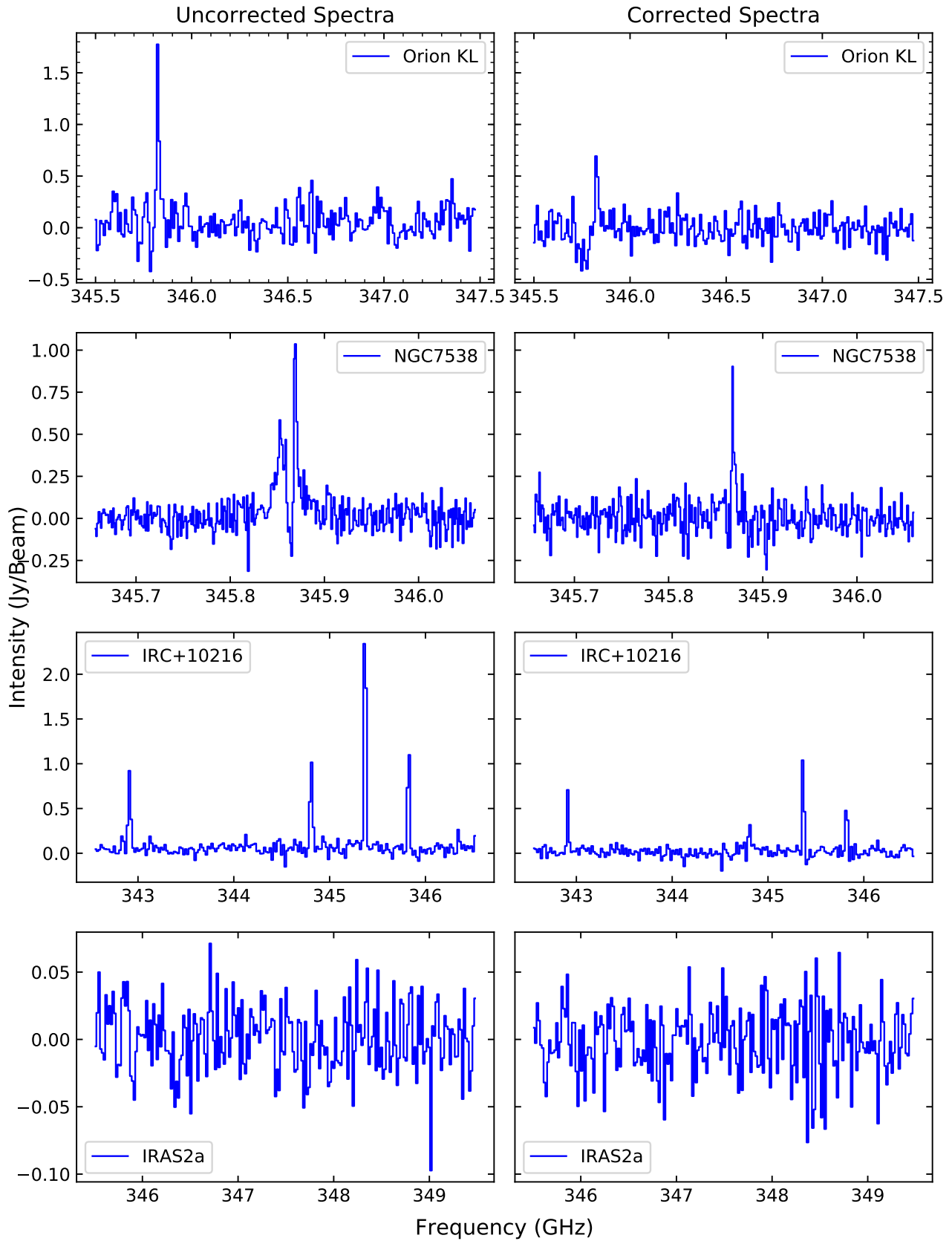}
\par\end{centering}
\caption{\label{fig:beforeafterspec}Stokes $V$ spectra of all objects before
and after squint correction.\emph{ Miriad}'s \texttt{maxfit} is used
on the CO map for each object to obtain the location in the image
where the Stokes $V$ signal at 345.8GHz is maximum, and a spectrum
is obtained through that point. Note that the Stokes $V$ signal decreases
in all cases after squint correction. The location of the peak varies:
sometimes it moves closer to the phase center after correction, sometimes
further away.}
\end{figure*}

Figures \ref{fig:results} and \ref{fig:iras2a} show corrected Stokes
$I$ and Stokes $V$ spectra (left) obtained at the peak of the CO
$(J=3\rightarrow2)$ Stokes $V$ signal on the corresponding maps
(right). A comparison of Stokes $V$ spectra before and after correction
are shown in Figure \ref{fig:beforeafterspec}. Notice in all cases
the Stokes $V$ signal decreases after squint correction. Stokes $V$
can also be found in the average spectrum obtained from all the visibility
data, though the SNR is of 3-5$\sigma$ significance in that case
compared to approximately 6-10$\sigma$ when the spectrum is taken
at the peak of the inverted maps. The presence of a Stokes $V$ signal
in the visibilities indicates the detections are not simply the result
of the inversion process that creates the maps. That is, the Stokes
$V$ signals are not due to sidelobes that appear when calculating
the inverse Fourier transform of the visibilities. As can be seen
in the maps, in general the peaks of Stokes $I$ and Stokes $V$ do
not coincide.

\begin{figure}
\begin{centering}
\includegraphics[bb=50bp 0bp 740bp 612bp,clip,scale=0.37]{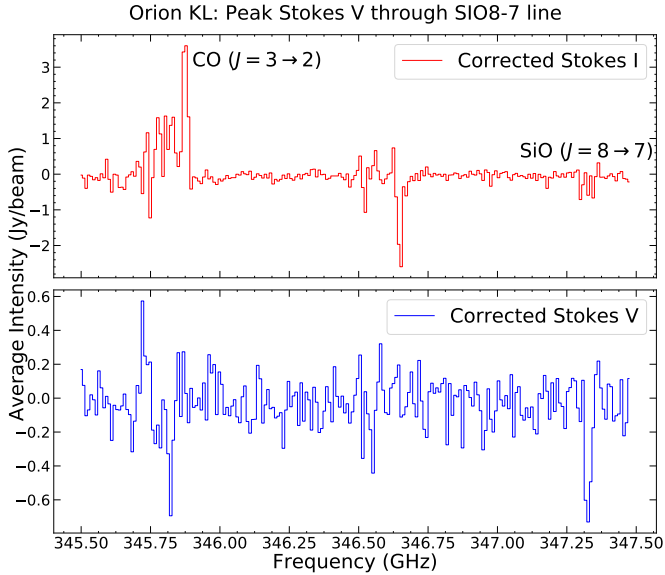}
\par\end{centering}
\caption{\label{fig:SiO}Peak Stokes $V$ signal for the SiO ($J=7\rightarrow8$
at 347.3 GHz) transition in Orion KL. Note there is also a strong
Stokes $V$ signal in the CO $(J=3\rightarrow2$ at 345.8 GHz) transition
here. The SiO signal is purely negative but the CO signal is antisymmetric.}
\end{figure}

In Orion KL (Fig. \ref{fig:results}) the lines of CO ($J=3\rightarrow2$
at 345.8 GHz) and SiO ($J=8\rightarrow7$ at 347.3 GHz) are both bright
(average visibilities show a peak Stokes $I$ of around 20 Jy/beam
and 55 Jy/beam, respectively) and both show Stokes $V$ signals. The
CO Stokes $V$ signal has an antisymmetric structure. On the other
hand, Figure \ref{fig:SiO} shows that the peak SiO Stokes $V$ signal
is purely negative. These lines (and others) are listed in Table \ref{tab:sumresults}
with their frequencies.

In IRC+10216 (Fig. \ref{fig:results}) we again see Stokes $V$ in
the CO ($J=3\rightarrow2$) but also in CS ($J=7\rightarrow6$), SiS
($J=19\rightarrow18$), and H$^{13}$CN ($J=4\rightarrow3$). 

In NGC7538 (Fig. \ref{fig:iras2a}) the Stokes $V$ signal in CO ($J=3\rightarrow2$)
at 345.8 GHz decreased in intensity after correction but is still
prominent. There was also initially a very strong Stokes $V$ signal
in CH$_{2}$CO at 346.6 GHz that completely disappeared after correction. 

Finally Figure \ref{fig:iras2a} shows no detection in NGC1333, with
only a clear detection of CO in Stokes $I$.

As mentioned previously, when assessing the Stokes $V$ detections
we consult the map for obvious pairs of positive/negative peaks that
would indicate beam offset and therefore a potentially false Stokes
$V$ signal. The maps are integrated over a narrow frequency band
of approximately 2 MHz so any peaks that exist are not washed out
by noise in adjacent channels. In the maps for Orion KL and IRC+10216
shown in Figure \ref{fig:results} there are no negative peaks around
the peak of Stokes $V$. However in NGC7538 there is quite a large
negative Stokes $V$ peak near our chosen peak that may indicate squint.
The top panel of Figure \ref{fig:beforeafter} shows a typical appearance
for squint signals which disappear after correction, and the pairs
tend to resemble each other in shape. The pair of peaks around our
chosen peak in NGC7538 however have distinct shapes. The worst case
here is that the signal is entirely squint but on the other hand the
signal may be a mixture of real and heavily affected by squint. The
detections in Orion KL and IRC+10216 are more reliable.

\begin{table}
\begin{centering}
\begin{tabular}{>{\raggedright}p{0.19\columnwidth}>{\raggedright}p{0.3\columnwidth}>{\raggedleft}p{0.1\columnwidth}>{\raggedleft}p{0.2\columnwidth}}
\toprule 
\textbf{Object} & \textbf{Line} & \textbf{(GHz)} & \textbf{Stokes $V$ (Jy/beam)}\tabularnewline
\midrule
\midrule 
\textbf{Orion KL} & CO $(J=3\rightarrow2)$ & 345.8 & 0.65\tabularnewline
\midrule 
 & SiO $(J=8\rightarrow7)$ & 347.3 & -0.65\tabularnewline
\midrule 
\textbf{NGC7538} & CO ($J=3\rightarrow2$) & 345.8 & 0.85\tabularnewline
\midrule 
\textbf{IRC+10216} & CS ($J=7\rightarrow6$) & 342.88 & 0.6\tabularnewline
\midrule 
 & SiS $(J=$19$\rightarrow$18) & 344.78 & 0.2\tabularnewline
\midrule 
 & H$^{13}$CN ($J=$4$\rightarrow$3) & 345.34 & 0.8\tabularnewline
\midrule 
 & CO ($J=3\rightarrow2$) & 345.8 & 0.4\tabularnewline
\midrule 
\textbf{NGC1333 (IRAS2a)} & CO ($J=3\rightarrow2$) & 345.8 & -\tabularnewline
\bottomrule
\end{tabular}
\par\end{centering}
\caption{\label{tab:sumresults}Summary of corrected Stokes $V$ signals found.
The beam size is determined by the configuration of the antennae array.
An intensity for the peak of the Stokes $V$ signal is only given
if the peak is noticeably higher than the noise level. The intensity
quoted for CO in NGC7538 is before smoothing is applied.}
\end{table}

\section{Discussion}

\label{sec:Discussion}The first question to address is whether our
CP detections are real or result from instrumental artifacts. This
is our chief concern because of the difficulty in calibrating CP measurements,
especially since the observations presented here were not made with
any special considerations for calibrating CP as in the observation
of Sgr A{*} reported in \citet{Munoz2012}. We will repeat here the
arguments made in Section \ref{sec:Observations} in support for the
soundness of these detections, and then discuss earlier detections
of CP and summarize how ARS can explain the detections presented in
this work.

Firstly, we take the average of all the visibility data (not shown)
and note that the peak Stokes $V$ is not proportional to the peak
Stokes $I$ at any particular frequency. For example, a large Stokes
$I$ at 347.25GHz (SiO ($J=8\rightarrow7)$) does not indicate a corresponding
peak in Stokes $V$. This property indicates that there is no systematic
leakage of Stokes $I$ into Stokes $V$, however excluding the longest
baselines from the averaging does cause the Stokes $V$ signals to
follow Stokes $I$, pointing to some instrumental CP in the shortest
baselines. 

In the average visibilities (not shown) of Orion KL CO ($J=3\rightarrow2$)
and SiO ($J=8\rightarrow7$) are the strongest lines, with the SiO
($J=8\rightarrow7$) line the stronger. Therefore, if the Stokes $V$
signal were purely leakage from Stokes $I$ then we would expect to
see an SiO Stokes $V$ signal that is stronger than the CO Stokes
$V$ signal in the visibilities, but we do not; the Stokes $V$ signal
from CO $(J=3\rightarrow2$) is stronger. This is also true in the
visibilities of IRC+10216, where the CS ($J=7\rightarrow6$) and SiS
$(J=19\rightarrow18)$ lines have similar strengths but the Stokes
$V$ at SiS $(J=19\rightarrow18)$ is twice as intense. However, in
the same object H$^{13}$CN $(J=4\rightarrow3$) and CO $(J=3\rightarrow2)$
have Stokes $V$ intensities that appear proportional to their Stokes
$I$ intensity (i.e., stronger $I$ means stronger $V$).

We also note that for Orion KL the shapes of the Stokes $V$ signals
vary across the frequency band and interpret this to mean that the
signals are likely not instrumental in nature, assuming that any instrumental
mechanism for producing spurious Stokes $V$ produces a single type
of CP (left or right). For example Figure \ref{fig:SiO} shows a spectrum
for that source with Stokes $V$ in CO and SiO. The SiO signal is
purely negative (indicating only LCP) but the CO signal is antisymmetric,
indicating the presence of both LCP and RCP. This has a physical explanation
using the ARS model in terms of blue-shifted and red-shifted scattering
populations that will be considered in Section \ref{subsec:ARS}.
We know of no instrumental mechanism for producing such a signature.
For the other objects the Stokes $V$ signal is always positive.

In the case of IRC+10216, an evolved carbon star with an extended
envelope, the peak of Stokes $V$ in the CO $(J=3\rightarrow2)$ map
(bottom-right panel of Figure \ref{fig:results}) is approximately
6'' away from the Stokes $I$ emission, which at first glance seems
to point to an erroneous detection. However the real size of the CO
envelope around IRC+10216 is much larger than seen here, as can be
verified from single-dish CO $(J=2\rightarrow1$) observations showing
a circumstellar shell with a radius of about 50'' \citep[Fig. 1 of][]{Cernicharo2015}.

Spatial filtering due to the resolution of the interferometer explains
the smaller spatial extent of IRC+10216 in the observations presented
here and also explains the frequent occurences of negative Stokes
$I$ in almost all the spectra shown in Figures \ref{fig:results},
\ref{fig:iras2a}, and \ref{fig:SiO}, as well as the high levels
of CP ranging from 6\% to 30\%. The largest resolvable scale by an
interferomter is determined by the length of the shortest baseline,
meaning that large scale emission can be filtered out by the interferometer.
For example in Orion KL the CO $(J=3\rightarrow2$) Stokes $I$ emission
is large and extended \citep{Hull2014}. If the Stokes $V$ signal
came from smaller localized areas, then we would observe peaks of
Stokes $V$ as shown while only a portion of the Stokes $I$ would
be present. The rest of the Stokes $I$ signal would be filtered away,
which could shift the zero-level downward. Fluctuations in Stokes
$I$ could then appear to have negative values, while the weaker Stokes
$I$ could also explain the high levels of $V/I$ observed.

Since the Stokes $V$ features shown in the maps of Orion KL and IRC+10216
are compact we made maps excluding the shortest baselines, which have
instrumental CP as shown from the averaging of the visibilities, and
included only the longest baselines. If they are real, we expect the
compact features to remain in these maps since the longest baselines
have the highest resolution. We saw that the Stokes $V$ peak remains
in the maps of IRC+10216 after excluding more than half of the baselines
as expected, however, the signal in Orion KL appears to disappear
after excluding the shortest half of the baseline distances. This
may indicate that the peak shown in Orion KL in Figure \ref{fig:results}
is instrumental.

The level of instrumental CP is in general greater the further the
peak is from the center of the map since the instrumental polarization
is found at the phase center. The peaks shown for Orion KL and NGC7538
(Figures \ref{fig:results} and \ref{fig:iras2a}) are several FWHMs
of the beam away from the center, meaning the level of instrumental
contamination in those peaks is likely higher.

Finally, we checked and confirmed that there were no Stokes $V$ signals
in the continuum comparable to those found in molecular lines like
CO and SiO. In the continuum, $V/I$ was found to be $<$1\% for Orion
KL and IRC+10216, and $<$4\% in NGC7538. If the detected Stokes $V$
originated from instrumental artifacts (other than $I$ leakage) then
we would expect to detect Stokes $V$ at similar levels in the continuum,
but this is not seen. The level of Stokes $V$ found in molecular
transitions is always higher than the level of Stokes $V$ in the
continuum for the observations presented here, and no significant
Stokes $V$ is found in the continuum.

We feel confident that the CP reported here, although perhaps suffering
from some level of instrumental contamination, is real and originates
from within these objects.

\subsection{Earlier Non-Zeeman CP Detections}

CP in a molecular spectral line weakly sensitive to the Zeeman effect
was first reported by \citet{Houde2013}, where approximately 2\%
polarization was detected in the $^{12}$CO $(J=2\rightarrow1)$ transition
at 230.5 GHz in Orion KL using the FSPPol at the CSO. The CP signal
was approximately symmetric (i.e. ``$\cap$''-shaped). The observation
was repeated three months after the first measurement to confirm the
result was not spurious, with similar results. Additionally the strong
line of HCN $(J=3\rightarrow2)$ at 265.9 GHz in Orion KL was measured
and no CP higher than the 0.1\% level was detected. The detection
in CO and the absence of a detection in HCN were evidence that the
FSPPol/CSO observations were not significantly suffering from leakage
into Stokes $V$ and highlighted the CO molecule as a target for non-Zeeman
CP. In all the objects presented here we find CP in $^{12}$CO $(J=3\rightarrow2)$
at 345.8 GHz (except for in NGC1333 IRAS2A where the CO line is relatively
weak). This is consistent with the original 2013 detection.

In follow up work \citet{Hezareh2013} examined the supernova remnant
IC 443 using dust polarimetry with PolKa at APEX and polarization
maps of $^{12}$CO $(J=2\rightarrow1)$ and $(J=1\rightarrow0)$ taken
with the IRAM 30m telescope. They initially found that the LP maps
of dust and CO differed greatly in their polarization angles. Expecting
that there was conversion of linear to circular polarization due to
ARS, the CO Stokes $V$ fluxes were then reinserted into the CO LP
signals. The resulting CO polarization angle maps then agreed very
well with each other, as well as with the dust map \citep[Fig. 9 of][]{Hezareh2013}.
This result clearly establishes a conversion from linear to circular
polarization.

The detections of CP in SiO ($\nu=1$ and $\nu=2$, $J=1\rightarrow0$)
masers at 43.1 and 42.8 GHz, respectively, were observed to have line
profiles inconsistent with the Zeeman effect \citep{Cotton2011}.
An attempt to explain the Stokes $V$ shapes with a non-Zeeman mechanism
involving the anisotropic pumping of the masers and a varying magnetic
field along the line of sight put forward by \citet{Wiebe1998} could
not account for the high levels of CP observed \citep{Cotton2011}.
We note that these SiO maser observations underwent a similar self-calibration
based process to correct for beam squint as the observations presented
here. 

\subsection{Anisotropic Resonant Scattering}

\label{subsec:ARS}ARS was the mechanism first proposed by \citet{Houde2013}
to explain the presence of CP in the transitions of CO. ARS rests
on a second-order interaction between radiation and matter in the
presence of a magnetic field. Incident radiation with photon states
polarized $\parallel$ and $\perp$ to the magnetic field can scatter
slightly differently off a molecule and incur a small phase shift
between the photon states. The phase shift incurred after propagating
and scattering off many molecules results in the appearance of CP
in the scattered radiation.

While this mechanism could reproduce the level of CP observed in CO
it initially failed to explain the observed symmetric ``$\cap$''-shaped
Stokes $V$ profile. In a follow up paper \citet{Houde2014} considered
the observations of Stokes $V$ in SiO masers \citep{Cotton2011}
and showed that the different profile shapes detected are readily
explained through ARS off populations of foreground molecules located
slightly outside of the velocity range of the line. For example, a
blue-shifted scattering population of molecules could result in a
negative ``$\cup$''-shaped profile while a red-shifted population
in a positive ``$\cap$''-shaped profile. The presence of both a
blue- and red-shifted population would cause an antisymmetric ``S''-shaped
profile (like the one seen in the top left panel of Figure \ref{fig:results}).

The conversion of LP to CP due to ARS can be illustrated by considering
background LP radiation oriented at some angle $\theta$ to the foreground
magnetic field. The incident and scattered radiation can be written
in terms of the \emph{n}-photon states as \citep{Houde2013}
\begin{eqnarray}
|\psi\rangle & = & \alpha|n_{||}\rangle+\beta|n_{\perp}\rangle\\
|\psi'\rangle & \simeq & \alpha e^{i\phi}|n_{||}\rangle+\beta|n_{\perp}\rangle,
\end{eqnarray}
where $\alpha=\cos(\theta)$, $\beta=\sin(\theta)$ and $\phi$ is
a phase shift incurred after multiple scattering events. Following
the definitions of the Stokes parameters and using an appropriate
basis the Stokes parameters for the scattered radiation can be found
to be 
\begin{eqnarray}
I & = & \alpha^{2}+\beta^{2}\\
Q & = & \alpha^{2}-\beta^{2}\\
U & = & 2\alpha\beta\cos(\phi)\\
V & = & 2\alpha\beta\sin(\phi).
\end{eqnarray}
This implies that in the chosen basis Stokes $U$ is lost to Stokes
$V$, i.e., more generally, LP is converted to CP. A calculation of
the phase shift $\phi$ incurred due to ARS can be found in \citet{Houde2013}.

Now, given a conversion from $U$ to $V$ it is clear that measuring
$V$ is necessary for techniques like the DCF method that rely on
the dispersion of the PAs of LP to calculate the strength of the magnetic
field. This is because without corresponding $V$ measurements, the
PAs obtained from a molecular spectral line subject to ARS (like CO
$(J=3\rightarrow2)$) will be rotated according to 
\begin{equation}
\tan(2\chi)=\cos(\phi)\tan(2\chi_{0}),
\end{equation}
where $\chi_{0}$ is the PA of the incident radiation and $\phi$
is the incurred phase shift as before \citep[eq. 10 of][]{Houde2014},
changing the dispersion of the PAs, as seen in \citet{Hezareh2013}.
As mentioned earlier, LP to CP conversion was reversed to obtain corrected
polarization angles. 

Because the DCF method relies on the observed dispersion of PAs using
LP, any such studies using molecular lines must also include corresponding
CP measurements to account for the polarization conversion effect
and determine the correct orientation of the PAs \citep{Hezareh2013,Chandrasekhar1953,Hildebrand2009,Houde2009}.

\section{Conclusion}

\label{sec:Conclusion}We analyzed polarimetric observations from
the SMA archive of Orion KL, IRC+10216, NGC7538 and NGC1333 IRAS2A
to search for CP signals. The data were corrected for squint, a source
of spurious Stokes $V$ signals that arises due to a slight misalignment
in the beams used to obtain Stokes $V$ when performing observations.
We found evidence of significant Stokes $V$ in Orion KL (in CO ($J=3\rightarrow2$)
and SiO $(J=8\rightarrow7$)), in IRC+10216 (in CS ($J=7\rightarrow6$),
SiS ($J=19\rightarrow18$), H$^{13}$CN ($J=4\rightarrow3$) and CO
($J=3\rightarrow2$)) and in NGC7538 (in CO ($J=3\rightarrow2$)).
We measured relatively important levels of CP ranging from 6\%-30\%,
probably due to the interferometric spatial filtering of large scale
emission. No significant Stokes $V$ was found in the continuum of
any of the objects.

Theories that explain the presence of non-Zeeman CP in molecular spectral
lines rely on the conversion of background LP to CP. The detections
in multiple lines and objects presented here indicate that such an
effect is likely widespread. Since the conversion of LP-to-CP modifies
the observed dispersion of PAs, it is necessary to obtain precise
measurements of CP along with LP for corresponding studies of the
magnetic field in the interstellar medium.

\acknowledgements{The Submillimeter Array is a joint project between the Smithsonian
Astrophysical Observatory and the Academia Sinica Institute of Astronomy
and Astrophysics and is funded by the Smithsonian Institution and
the Academia Sinica. M.H. is supported by the Natural Science and
Engineering Research Council of Canada Discovery Grant RGPIN-2016-04460.
J.M.G is supported by the MINECO (Spain) AYA2014-57369-C3 and AYA2017-84390-C2
grants.\\
}

Reduction scripts are available at \url{http://github.com/mef51/SMAData}.
A Python wrapper for \emph{Miriad} including the squint correction
script used is available at \url{http://github.com/mef51/smautils}.

\end{document}